\documentclass[twocolumn,aps,prc,qshowpacs,floatfix,nofootinbib,superscriptaddress]{revtex4-1}
\usepackage{amsmath}
\usepackage{graphicx}
\usepackage{url}
\usepackage[table]{xcolor}
\usepackage{multirow}
\usepackage{longtable}
\usepackage{color}
\usepackage{hyperref}
\usepackage[normalem]{ulem}  
\usepackage{float}
\usepackage{amsmath,amssymb}
\usepackage[caption = false]{subfig}

\begin{document}
\title{Formation of light nuclei at chemical freezeout: Description within a statistical thermal model}
\author{Deeptak Biswas}
\email{deeptak@jcbose.ac.in}

\affiliation{
Department of Physics, Center for Astroparticle Physics \& Space Science\\
Bose Institute, EN-80, Sector-5, Bidhan Nagar, Kolkata-700091, India 
}
%
%
\begin{abstract}

The thermal description of light nuclei at the chemical freeze-out has been investigated. First, I have verified the equilibration of the light nuclei and then introduced a new method to investigate the light nuclei formation. One can study the proximity between the phase space density of light nuclei ratios and their hadronic constituents e.g $\bar{d}/d$ and $(\bar{p}\bar{n}/pn)$. I have found that with the exclusion of the decay feed-down from the hadronic yields in the thermal model, the hadronic representations have good agreement with the light nuclei ratios. I have performed a similar analysis with the ratio of $\Lambda$-hypernuclei and $^3 He$, which is related to the ratio $\Lambda/p$. In this context, the strangeness population factor $S_3$ has been studied also. These results indicate that the nuclei and hypernuclei formation may occur near the standard chemical freeze-out and before the decay of the hadronic resonances. This method will serve as a guideline to discuss the light nuclei formation and the inclusion of decay into their hadronic constituents.

\end{abstract}

\maketitle

\section{\label{sec:Intro} Introduction}
The light nuclei and hypernuclei yields are available for a  wide range of  collision energies, from AGS \cite{Ahle:1999in, Armstrong:2002xh}, SPS \cite{Anticic:2016ckv} to  RHIC \cite{Abelev:2010rv, Adam:2019wnb} and LHC \cite{Adam:2015vda, Adam:2015yta, Acharya:2017bso}. The existence of these light nuclei at the chemical freeze-out boundary is uncertain, as their binding energies (few MeV) are much lower than the typical freeze-out temperature ($150~\mathrm{MeV}$)\cite{Andronic:2017pug}. Despite these difficulties, a thermal model representation of these bound states is important to understand the degree of equilibration of the produced fireball. The formation of light nuclei is also crucial in the cosmological context. As an example, the generated deuterons could be dissociated into their constituent nucleons if produced in an earlier epoch. Their production could be favorable only when photon decoupled from baryons and the process $n + p \rightarrow d + \gamma$ became dominant in the detailed balance \cite{Braun-Munzinger:2018hat}.

Statistical Hadronization Model (SHM) is a standard prescription to discuss the hadronic yields of heavy-ion collision. This formalism is quite successful in explaining the final abundances of hadrons, with only a  limited number of thermodynamic parameters (T, $\mu_B$, $\mu_Q$, $\mu_S$, V) \cite{BraunMunzinger:1994iq, Andronic:2005yp, Andronic:2008gu, Andronic:2010qu, Andronic:2011yq, Andronic:2012ut, Andronic:2017pug}. The surface of these parameters is known as the Chemical Freeze-out (CFO), as inelastic collision terminates and the $p_T$ integrated hadron yields are frozen onward this boundary. The contradiction arises while describing the light nuclei in this framework of this thermal model. These nuclei should not survive the chemical freeze-out due to their smaller binding energy, and collisions with pions will dissociate these nuclei into constituent nucleons \cite{Oliinychenko:2018ugs}. Recently Ref.\cite{Cai:2019jtk} has also discussed the assumptions of the SHM and raised concerns about the yields of these weakly bound states at the chemical freeze-out surface. 

The Coalescence model also addresses the hadrons formation of heavy-ion collisions \cite{Scheibl:1998tk, Fries:2003vb, Hwa:2003bn, Molnar:2003ff}. In this model, depending on the momentum and spatial distribution, nearby partons confine to form a hadron. At the phenomenological level, this method relies on the momentum spectra of both the constituents and the final bound state. A complete description of local correlation and energy conservation is not possible due to the absence of experimental measurement of the parton spectra. On the other hand, the discussion of the light nuclei formation is simpler as the measured momentum spectra are available for both the light nuclei and their hadronic constituents \cite{Abelev:2009ae}. Two or more hadrons coalesce to form the light nuclei near the kinetic freeze-out surface. The momentum spectra of a light nuclei with $Z$ protons and $A-Z$ number of neutrons is proportional to, $\left(E_{p} \frac{d N_{p}}{d^{3} p}\right)^{Z}\left(E_{n} \frac{d N_{n}}{d^{3} p}\right)^{A-Z}$. This method has to implement several parameters to discuss the experimental data. One can calculate the hadron yields and their ratios from the thermodynamic description of the chemical freeze-out. As the nucleons further coalesce to form light nuclei, a one to one mapping in chemical composition between the light nuclei and their constituents is apparent.

Despite these variations, both thermal and coalescence models make similar predictions of light nuclei yields \cite{BraunMunzinger:1994iq, Mrowczynski:2016xqm}. These light nuclei and hypernuclei, especially (anti-)deuterons are cleaner probes of the chemical freeze-out for having a negligible decay contribution from the higher mass clusters \cite{DeMartini:2020hka, Vovchenko:2020dmv, Oliinychenko:2020ply}. So from a parametrization of the statistical thermal models, one can directly calculate the yields of these nuclei and compare it with the experimental data.

Ref.\cite{Cleymans:2011pe} analyzed the ratio of light nuclei and their constituents, assuming the Boltzmann approximation and neglecting the decay feed-downs into hadrons. Though the deuteron to proton ratio was successfully reproduced in this method, the hypernuclei to light nuclei ratio did not agree with the data. With hypernuclei data from RHIC-200 GeV, it remains a challenge for thermal models to simultaneously describe all hadrons and hypernuclei in a single freeze-out picture. Ref.\cite{Chatterjee:2014ysa} utilized two separate freeze-out surfaces for strange and non-strange particles to address this issue. Recently, ref.\cite{Oliinychenko:2018ugs} has shown identical production and disintegration rates for deuterons in a hydrodynamical approach, which holds even in the presence of baryon-antibaryon annihilation.

The $\Lambda$ hypernuclei production is related to the primordial $\Lambda$-p phase space correlation. Referring to this, the strangeness population factor $S_{3}={_{\Lambda}^{3}\mathrm{H}}/\left(^{3} \mathrm{He} \times \frac{\Lambda}{p}\right)$ was proposed \cite{Zhang:2009ba}. A multiphase transport model (AMPT) shows an enhancement of this ratio in case of a deconfined initial state, relative to a system with only a hadronic phase. This ratio is also important to investigate the strangeness baryon correlation $C_{BS}$. 

The present work reviews the thermodynamics of the chemical freeze-out and considers a uniform thermal description for the hadrons and light nuclei. I have verified the equilibration of the light nuclei in this prescription and also investigate the ratios concerning the hypernuclei and strangeness population factor $S_3$. The resulted parametrization has reasonably reproduced $S_3$ at RHIC-200 GeV and LHC-2760 GeV. As these weakly bound states are composed of hadrons, so one can ask, whether these light nuclei formation happens near the hadronic chemical freeze-out or some later times, and do these light nuclei experience a similar chemical freeze-out?  In a thermal model, the inclusion of resonance decay may help to investigate these questions regarding the light nuclei formation and freeze-out.

One can represent the light nuclei ratios with their hadronic constituents e.g the ratio $\bar{d}/d$ can be approximated with $(\bar{p}/p)^2$. If the light nuclei are produced near the chemical freeze-out boundary and immediately experience the freeze-out, then a hadronic description with only the primary yields of hadrons should be a reasonable representation for the phase space distribution of these nuclei and hypernuclei ratios. Whereas, if the hadrons produce these bound states long after the chemical freeze-out, then decay feed-downs from higher mass resonance will be added to the final yields of the hadrons. On this occasion, the light nuclei ratios will have a better resemblance to the ratio of total yields (primary plus decay feed-downs) of the hadronic constituents. I have tried to address these issues in the present manuscript. Though the parametrization has been performed with the proper decay contribution into final hadron states, one can find that the hadronic description provides a better estimation for the light nuclei ratios with the exclusions of the feed-down of higher mass resonances. This study suggests that the light nuclei yields attain an equilibrium value near chemical freeze-out, and this formation of nuclei and hypernuclei occurs long before the decay feed-down to nucleons and hyperons take place. 
 
The manuscript is organized as follows. Section~\ref{sec2} will discuss the parametrization procedure and introduce essentials tools to discuss the findings. In section~\ref{sec3} I shall discuss the results and summarize in section~\ref{sec4}.
 
\section{Formalism}\label{sec2}
In this section, I shall briefly discuss the parameterization method and available experimental data of the light nuclei sector.
\subsection{Parameterization with hadron resonance gas}
The ideal hadron resonance gas is an effective tool to describe the matter at freeze-out. For the last two decades, several studies have successfully explained the bulk properties of heavy ion collision at freeze-out by applying this model \cite{BraunMunzinger:1994xr, Cleymans:1996cd, BraunMunzinger:1999qy, Cleymans:1999st, BraunMunzinger:2001ip, Becattini:2005xt, Andronic:2005yp, Andronic:2008gu}. At the chemical freeze-out, one can associate particle density with experimentally measured yield by, \cite{Manninen:2008mg},
\begin{equation}{\label{eq.solve}}
\frac{dN_i}{dy}|_{Det}={\frac{dV}{dy}}n_i^{Tot}|_{Det}
\end{equation}
where the subscript $Det$ denotes the detected hadrons. The total number density of any hadron is,
\begin{eqnarray}
&n_i^{total}& ~=~ n_i^{primary}(T,\mu_B,\mu_Q,\mu_S)~ + 
\nonumber \\
&\sum_j& n_j(T,\mu_B,\mu_Q,\mu_S) \times \mathrm{Branching~Ratio} (j
\rightarrow i)
\end{eqnarray}
where the summation runs over the heavier resonances ($j$), which decay to the
$i^{th}$ hadron and $primary$ denotes the thermal density of hadrons without decay contribution. 

The number density $n_i$ is calculated using Eq.\ref{density}.
\begin{equation}{\label{density}}
 n_i =\frac{T}{V} \left(\frac{\partial \ln Z_i}
       {\partial\mu_i}\right)_{V,T} 
 =\frac{g_i}{{(2\pi)}^3} \int\frac{d^3p} {\exp[(E_i-\mu_i)/T]\pm1}.
\end{equation}
For the $i^{th}$ species of hadron, $g_i$, $E_i$ and $m_i$ are respectively the degeneracy factor, energy, and mass, whereas $\mu_i=B_i\mu_B+S_i\mu_S+Q_i\mu_Q$ is the chemical potential, with $B_i$, $S_i$ and $Q_i$ denoting the baryon number, strangeness and the electric charge respectively. Though this model is commonly applied for hadrons and their resonances, I can incorporate the light nuclei states with their respective quantum numbers, mass, and degeneracy \cite{Andronic:2010qu, Andronic:2011yq, Andronic:2017pug}. 

Here I have followed a recently introduced formalism for the chemical freeze-out parameter extraction \cite{Bhattacharyya:2019wag, Biswas:2020dsc}. This approach relies on ratios of conserved current like net baryon charge and entropy and suitably parameterizes the freeze-out surface with good precision. In this method, one constructs net charges and total charges from the detected particle's rapidity spectra and equates the model estimation of the net baryon number normalized to the total baryon number with that of the experimental data, as in Eq.\ref{eq.conserveb}. The other equation is constructed for detected net baryon number normalized to total particle yield as Eq.\ref{eq.t}. 
\begin{eqnarray}
\frac{\sum_i^{Det} B_i \frac{dN_i}{dY}}{\sum_i^{Det} |B_i|
\frac{dN_i}{dY}}
&=& \frac{\sum_i^{Det} B_i n_i^{Tot}}{\sum_i^{Det} |B_i| n_i^{Tot}} 
\label{eq.conserveb} \\
\frac{\sum_i^{Det} B_i \frac{dN_i}{dY}}
{\sum_i^{Det}\frac{dN_i}{dY}}
&=& \frac{\sum_i^{Det} B_i
n_i^{Tot}}{\sum_i^{Det} n_i^{Tot}} 
\label{eq.t}
\end{eqnarray}
The last equation relies on the fact that the detected total particle multiplicity is a good measure of total entropy \cite{Biswas:2019wtp}. These two equations are solved alongside two constraints of the colliding nuclei, i.e net electric charge to net baryon and strangeness neutrality. The systematics regarding the freeze-out volume is nullified as one deals with ratios only. It is important to note, though this formalism is different from the standard $\chi^2$ analysis, the extracted parameter set is consistent with the results from Ref.\cite{Andronic:2005yp, Becattini:2005xt, Adamczyk:2017iwn, Bhattacharyya:2019wag, Biswas:2020dsc}.

Here it should be mentioned that the yields of these light nuclei are considerably smaller than that of the hadrons. So, the addition of light nuclei in the parametrization process should not significantly affect the extracted parameters. 

\subsection{Data analysis}
I have used SPS data of deuterons, $\mathrm{^3 He}$, and $\mathrm{^3 H}$ following Ref.\cite{Anticic:2016ckv} and data for Au-Au collision in BES from Ref.\cite{Adam:2019wnb}. Only in LHC, the data of all four light nucleus are available ($d$, $^3 \mathrm{He}$, $^3_\Lambda \mathrm{H}$, and $^4 \mathrm{He}$) \cite{Adam:2015vda, Adam:2015yta, Acharya:2017bso}. I have included AGS data of 11.6 AGeV/c beam energy for proton and deuteron from  Ref.\cite{Ahle:1999in}. Ratios regarding hypertriton($^3_\Lambda \mathrm{H}$), $^3 \mathrm{H}$ are given in Ref.\cite{Abelev:2010rv} for RHIC 200 GeV. As individual yields are not available, I could not utilize most of these yields in this analysis except at LHC energy. Ratios have been predicted from the resulted parametrization and compared with available data.  

I have included mid-rapidity yields ($dN/dy$) of hadrons for most central collision following AGS~\cite{Ahle:1999uy, Ahle:2000wq, Klay:2003zf, Klay:2001tf, Back:2001ai, Blume:2011sb, Back:2000ru, Barrette:1999ry, Back:2003rw}, SPS~\cite{Alt:2007aa, Alt:2005gr, Afanasiev:2002mx, Afanasev:2000uu, Bearden:2002ib, Anticic:2003ux, Antinori:2004ee, Antinori:2006ij, Alt:2008qm, Alt:2008iv, Anticic:2003ux}, RHIC~\cite{Kumar:2012fb, Das:2012yq, Adler:2002uv, Adams:2003fy, Zhu:2012ph, Zhao:2014mva, Kumar:2014tca, Das:2014kja, Abelev:2008ab, Aggarwal:2010ig, Abelev:2008aa, Adcox:2002au, Adams:2003fy, Adler:2002xv, Adams:2006ke, Adams:2004ux, Kumar:2012np, Adams:2006ke} and LHC~\cite{Abelev:2012wca, Abelev:2013xaa, ABELEV:2013zaa, Abelev:2013vea}. Data for STAR BES has been used following~\cite{Adamczyk:2017iwn, Adam:2019koz}. In the considered HRG spectrum, all confirmed hadronic states up to mass 2 GeV have been included, with masses and branching ratios following the Particle Data Group~\cite{Tanabashi:2018oca} and THERMUS \cite{Wheaton:2004qb}. Finally, I have solved Eq.(\ref{eq.conserveb}$-$\ref{eq.t}) and the constraints numerically, using Broyden's method with a minimum convergence criterion of $10^{-6}$. The variances of thermal parameters have been estimated by repeating the analysis at the given extremum value of hadrons yields. The errors in experimental data points are the quadrature sum of statistical and systematic uncertainties.

\section{Result and discussion}\label{sec3}
The extracted parameter set $(T,\mu_B,\mu_Q,\mu_S)$ has good agreement with previous analyses, which were obtained with only hadron yields \cite{Bhattacharyya:2019wag, Bhattacharyya:2019cer, Biswas:2020dsc, Biswas:2020opf}. At the LHC energy, the temperature decreases $1~\mathrm{MeV}$ if one incorporates all available light nuclei yields. This variation is within the estimated variances of Ref.\cite{Bhattacharyya:2019wag, Bhattacharyya:2019cer, Biswas:2020dsc}.
\begin{figure}[!htb]
{\includegraphics[scale=0.85]{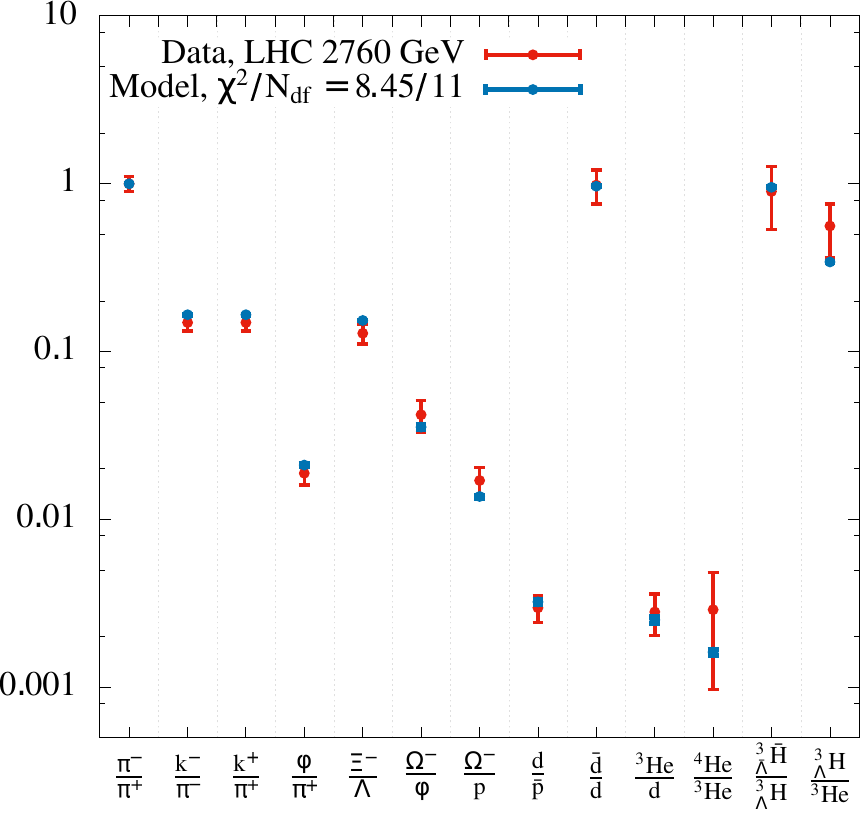}}
\caption{
\label{fg.2760}
(color online) Ratio of various particles and light nuclei yields for LHC 2.76 TeV. Data (Red line) are from Ref.\cite{Abelev:2012wca, Abelev:2013xaa, ABELEV:2013zaa, Abelev:2013vea, Adam:2015vda, Adam:2015yta, Acharya:2017bso}. Blue lines are thermal prediction. Estimated $\chi^2$ by degrees of freedom for the representative ratios is 8.45/11.}
\end{figure}

It is a general exercise to reproduce particle ratios with the extracted parameter set to verify the accuracy of the fitting procedure. I have used all the available light nuclei yields in the fitting for the LHC energy. The predicted ratios regarding meson, baryon, and light nuclei, alongside their experimental data, are shown in Fig.[\ref{fg.2760}]. All the particle ratios have been reproduced with excellent precision. One should consider that this parameterization method does not depend on individual yield ratio, so these ratios are independent predictions. The particle and anti-particle yields become identical at LHC, which demands the chemical potentials to be zero. The resemblance between $k^+/\pi^+$ and $k^-/\pi^-$ is also an indication of the vanishing $\mu_S$. The agreement between data and thermal model prediction establishes the fact that the light nuclei and hadrons experience the same chemical freeze-out. This fact raises contradictions due to the smaller binding energy of light nuclei. The light nuclei should melt immediately at a freeze-out temperature of $152$ MeV. Despite this discrepancy, the beautiful agreement at LHC makes it interesting to investigate ratios regarding light nuclei at the other collision energies.
\subsection{Light nuclei to proton ratio}
\begin{figure}[!htb]
{\includegraphics[scale=0.85]{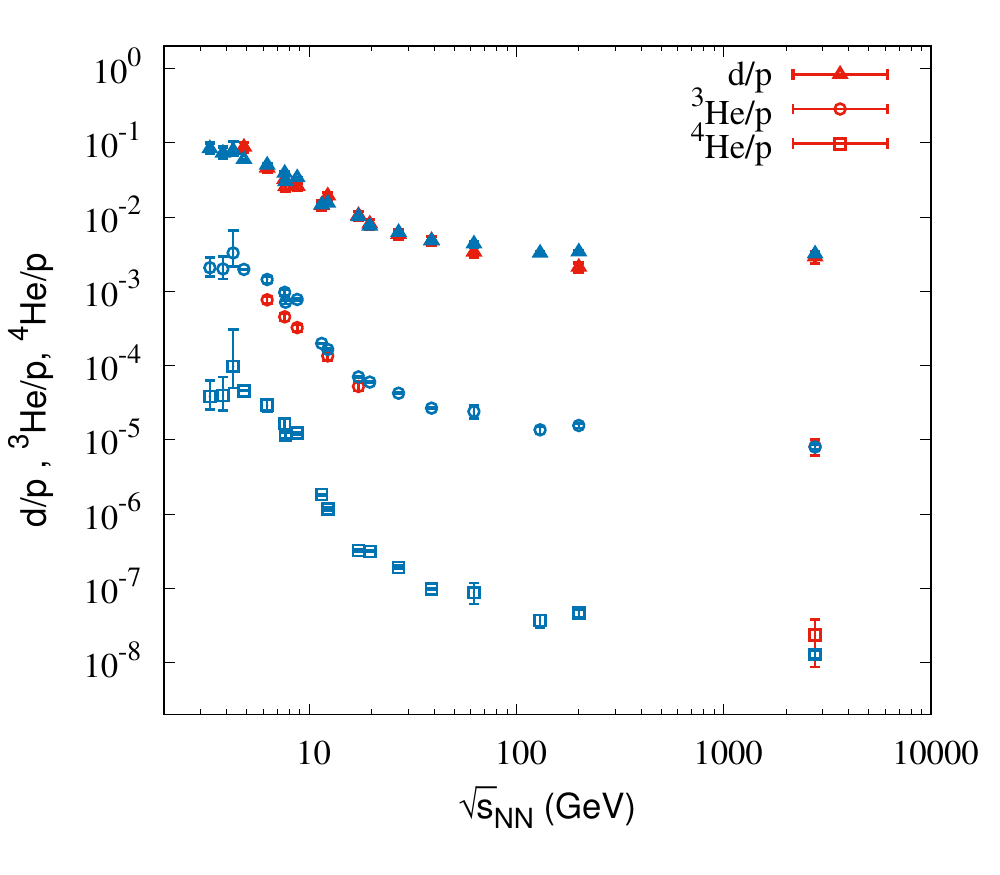}}
\caption{
\label{fg.nucleitop}
(Color online) Variations of light nuclei to proton ratio with $\sqrt{s_{NN}}$. The red points are the data from AGS\cite{Ahle:1999in} SPS\cite{Anticic:2016ckv}, RHIC \cite{Adam:2019wnb} and LHC \cite{Adam:2015vda, Acharya:2017bso} . The blue points are the model predictions.}
\end{figure}

Light nuclei yields are significant to review the baryon equilibrium for their high baryon content. One can normalize the light nuclei (d, $^3\mathrm{He}$, $^4\mathrm{He}$ ) yields with proton yields to examine the collision energy variation, as presented in Fig.[\ref{fg.nucleitop}]. Measured yields of the deuteron are available from RHIC-BES, LHC, whereas estimations for $^3\mathrm{He}$ are available at SPS and LHC. There is reasonable agreement between the model predictions and experimental data, which indicates the chemical equilibrium of these light nuclei states at freeze-out.

These three ratios show a similar variation with the collision energy ($\sqrt{s_{NN}}$). They remain flat at the higher RHIC, LHC, and increase towards lower BES and AGS energies. The relative difference between LHC and AGS values increases with the mass number of light nuclei \footnote{Two orders of magnitude for $d/p$, whereas $^4\mathrm{He}/p$ rises to $10^{-4}$ in AGS, from $10^{-8}$ of LHC energy}. At the lower collision energies, a finite $\mu_B$ favors the production of baryon clusters with a higher baryon number. Whereas, at the higher RHIC and LHC, the light nuclei yields are just mass suppressed. This explains the variation shown. From the parametrization, one can observe a horn in the $^3 \mathrm{{He}/p}$ and $^4 \mathrm{{He}/p}$ at lower AGS energy. This peak arises as an interplay among the thermal parameters and nucleon mass. Future data from CBM and NICA collaborations will help to investigate these claims. 

\subsection{Anti-particle to particle ratio of d and p }
\begin{figure}[!htb]
\subfloat[]{
{\includegraphics[scale=0.85]{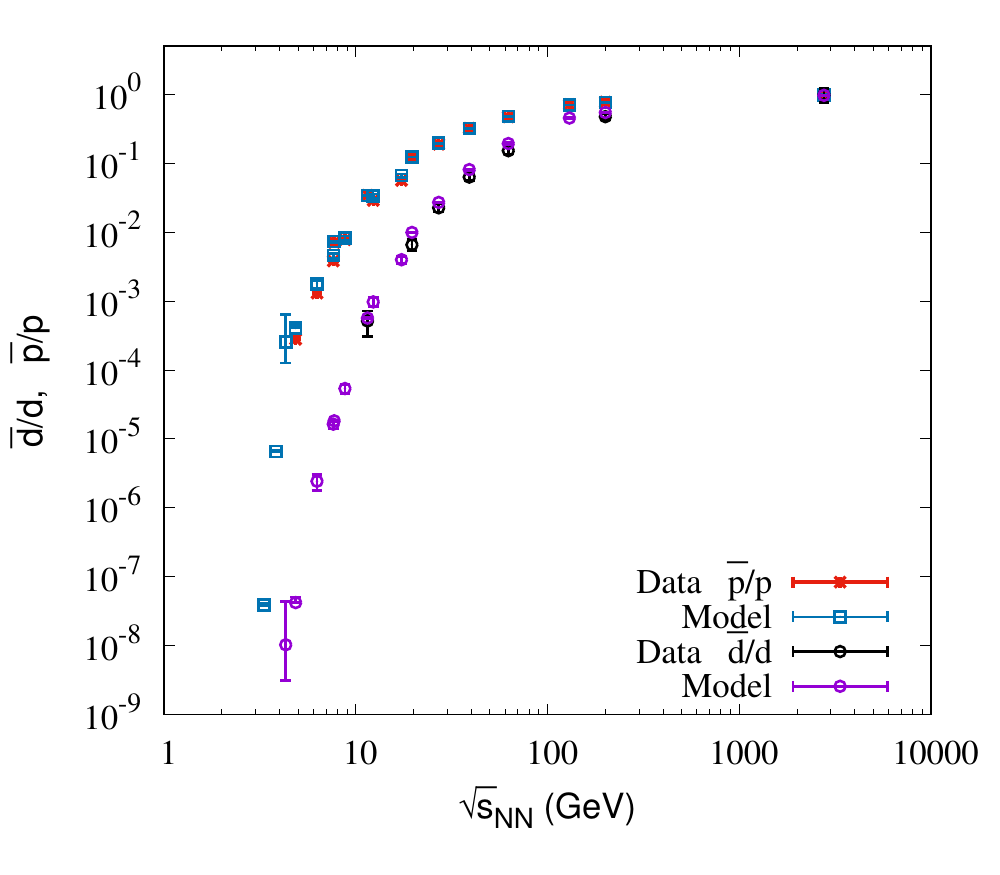}}
\label{fg.dueteron}
}\\
\subfloat[]{
{\includegraphics[scale=0.85]{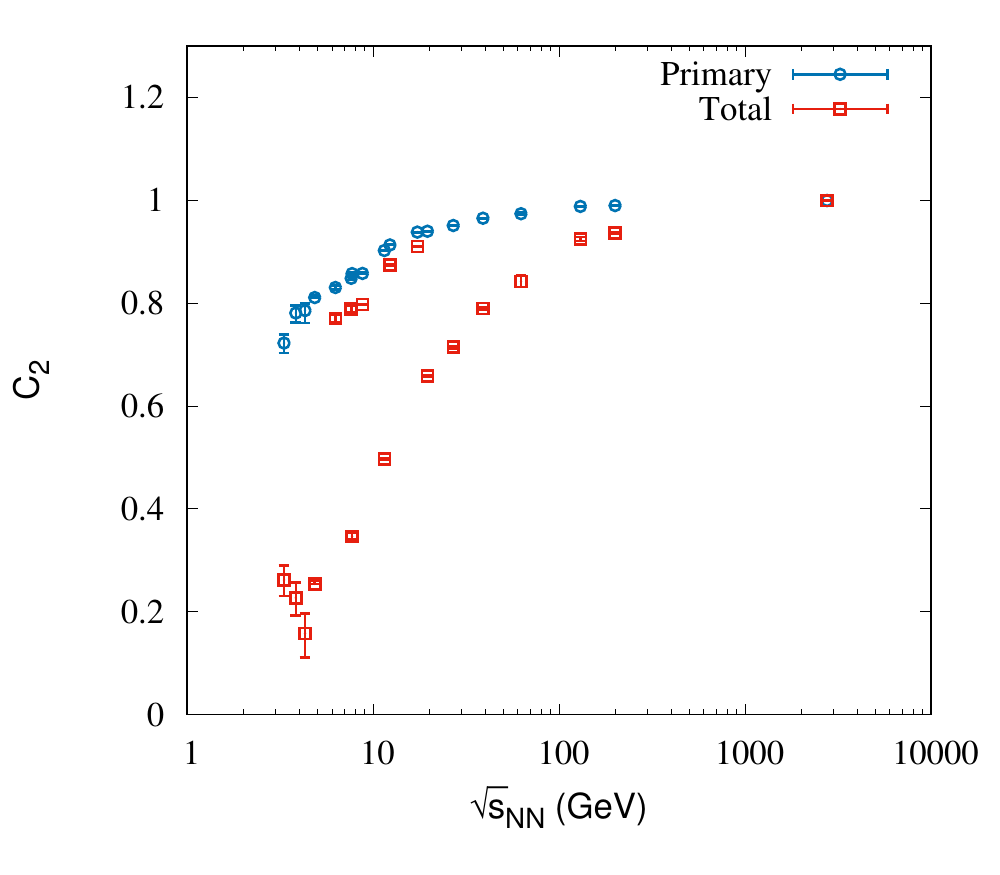}}
\label{fg.dfromp}
}
\caption{
\label{fg.ddbar}
(a) Red and black points are data \cite{Adam:2015vda, Adam:2019wnb} for $\bar{p}/p$ and $\bar{d}/d$ respectively. Blue and violet points denote model estimations. (b) Variation of $C_2$  with $\sqrt{s_{NN}}$. The red and blue points denote estimations with and without decay feed-down into (anti-)proton yield respectively.}
\end{figure}

Fig.[\ref{fg.dueteron}] presents the antiproton to proton and anti-deuteron to deuteron ratio. The model estimations suitably match with the experimental data. Both of these ratios increase with the collision energy and become $1$ at LHC, as the particle and antiparticle yields become equal. On the other hand, due to a large baryon stopping among the colliding nuclei (which results in a finite $\mu_B$), the baryons are more abundant than the anti-baryon at lower $\sqrt{s_{NN}}$. This demands $\bar{d}/d$ to be smaller than $\bar{p}/p$, as deuteron has a larger baryon content. The agreement of the thermal model with data elucidates the existence of (anti-)deuterons at the hadronic chemical freeze-out. 

The decay contribution from the higher mass clusters into (anti-)deuteron is negligible \cite{DeMartini:2020hka, Vovchenko:2020dmv, Oliinychenko:2020ply}, so these yields can be determined directly from the primary thermal density. The (anti-)deuteron is a weakly bound state of neutron and proton. In a general coalescence picture, the light nuclei density is proportional to their constituents' thermal abundances \cite{Scheibl:1998tk, Cleymans:2011pe, Chatterjee:2014ysa}. Neglecting the isospin asymmetry, one can assume proton and neutron density to be equal and approximates $\mathrm{\bar{d}/d}$ with the squared anti-proton to proton ratio \cite{Cleymans:2011pe}.

\begin{equation}
\frac{\bar{d}}{d}= C_2 \left(\frac{\bar{p}\bar{n}}{pn}\right) \simeq C_2 \left(\frac{\bar{p}}{p}\right)^2
\end{equation}

This $C_2$ helps to investigate the light nuclei formation by quantifying the similarity in chemical composition between $\bar{d}/d$ and $\mathrm{(\bar{p}/p)^2}$. To do that, one can consider the hadronic ratios from thermal parametrization in two scenarios. First, I have estimated $\mathrm{(\bar{p}/p)^2}$ with only $primary$ yields of (anti-)proton. A better resemblance of $\bar{d}/d$ with this $primary$ $\mathrm{(\bar{p}/p)^2}$ will imply that the (anti-)deuteron formation happens from the primordial (anti-)protons. In the second case, $\mathrm{(\bar{p}/p)^2}$ has been constructed including the decay feed-down in the (anti-)proton yields. If the (anti-)deuterons are formed long after the chemical freeze-out, then the square of this total antiproton-proton ratio will be a good representation for $\bar{d}/d$.

In Fig.[\ref{fg.dfromp}] the collision energy variation of $C_2$ has been presented, for both the cases. $C_2$ increases with $\sqrt{s_{NN}}$ and saturates near $1$ at RHIC and LHC. This variation is comparatively smaller ($0.8$ to $1$) if one evaluates $(\bar{p}/p)^2$ entirely from primary density of the (anti-)proton. On the contrary, $C_2$ decreases significantly in lower $\sqrt{s_{NN}}$ with the inclusion of resonance decay. In lower AGS and BES energies, the feed-down contributions from the baryonic resonances are larger than anti-baryons due to the finite $\mu_B$, which increases the asymmetry between the total yields of proton and antiproton. The higher value of $C_2$ for the $primary$ case denotes that $(\bar{p}/p)^2$ with the primordial yields of (anti-)proton is a better representation of $\bar{d}/d$. In this case, the little deviation from 1 at lower collision energy can be reduced by considering the isospin asymmetry and neutron yields properly. This finding means that the (anti-)deuterons are formed from the primary (anti-)nucleons, near the chemical freeze-out boundary. As I have already presented a good agreement between the thermal model and experimental data for both the ratios, this finding will act as a benchmark to study the light nuclei formation.

This conclusion is in agreement with the findings of ref.\cite{Oliinychenko:2018ugs}. They have observed that the deuteron yields become fixed near the chemical freeze-out, though the inelastic interactions may continue further. Here one should consider, that the yields of baryon and antibaryon are equal at LHC, so the antiproton to proton ratio does not vary with the inclusion of feed-down. The hypernuclei to light nuclei ratios will be relevant in this context 

\subsection{Hypertriton to $^3 \mathrm{He}$ ratio}
Hypernuclei are produced in high-energy interactions via hyperon capture by nuclei \cite{Botvina:2016wko}. The lowest mass hypernuclei are $\Lambda$-hypertriton (${^3_\Lambda}\mathrm{H}$). In a thermal model, yields and ratios regarding this hypernuclei support to understand the phase space occupancy for strangeness at the freeze-out. For example, a hypertriton is a bound state of n, p, and $\Lambda$. On the other hand, ${^3}\mathrm{He}$ has two protons and one neutron. This resemblance of these two states makes their ratio important for investigating strangeness equilibration. In a coalescence picture, the ratio ${^3_\Lambda}\mathrm{H}/{^3}\mathrm{He}$ should follow the $\Lambda/p$ ratio. A ratio $S_3$, namely the strangeness population factor has been proposed \cite{Zhang:2009ba}, where
\begin{equation}
\left(\frac{^3_\Lambda \mathrm{H}}{^3 \mathrm{He}}\right)= \left(\frac{\Lambda \mathrm{p n}}{\mathrm{ ppn}}\right) = S_3  \left(\frac{\Lambda}{\mathrm{p}}\right) 
\end{equation}

and 

\begin{equation}
S_3 = \left(\frac{^3_\Lambda \mathrm{H}}{^3 \mathrm{He}}\right) / \left(\frac{\Lambda}{\mathrm{p}}\right) 
\end{equation}

\begin{figure}[!htb]
\subfloat[]{
{\includegraphics[scale=0.85]{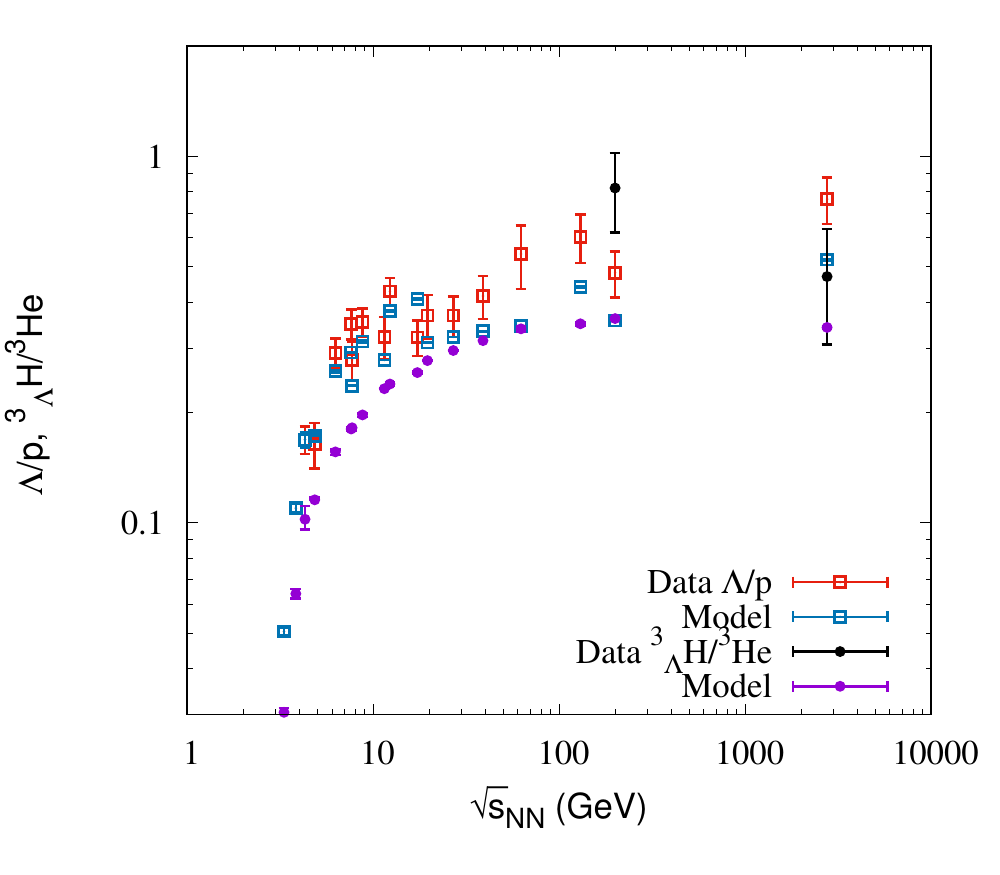}}
\label{fg.lambdap}
}\\
\subfloat[]{
{\includegraphics[scale=0.85]{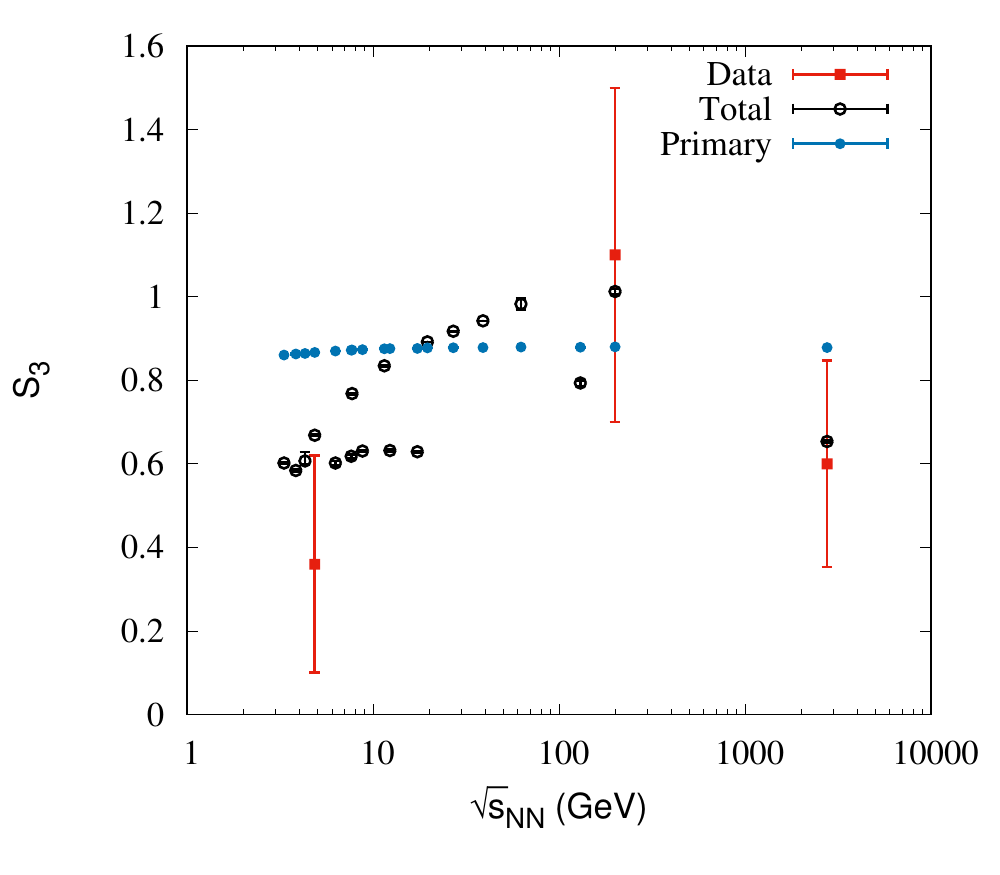}}
\label{fg.S3}
}
\caption{
\label{fg.hyperontoH}
(a) Collision energy variation of $\Lambda/p$ and ${^3_\Lambda}\mathrm{H}/{^3}\mathrm{He}$. The ratio regarding hypernuclei are only available in LHC \cite{Adam:2015yta} and RHIC 200 Gev \cite{Abelev:2010rv}. Red and black are the data points. Blue and violet points are the model predictions for $\Lambda/p$ and ${^3_\Lambda}\mathrm{H}/{^3}\mathrm{He}$ respectively. (b) Variation of $S_3$  with $\sqrt{s_{NN}}$. Here black points are estimations with total $\Lambda$, proton yield and blue denotes $S_3$ without the decay feed-down. Red points are the experimental data regarding $S_3$. AGS Data are from Ref.\cite{Armstrong:2002xh}.}
\end{figure}

In Fig.[\ref{fg.lambdap}], the ratios $\Lambda/p$ and ${^3_\Lambda}\mathrm{H}/{^3}\mathrm{He}$ have been displayed. Data are only available at LHC \cite{Adam:2015yta} and RHIC 200 Gev \cite{Abelev:2010rv} for the hypernuclei to nuclei ratio. The predicted $\Lambda/p$ has good agreement with experimental data. In RHIC energies, the difference between data and model prediction is an influence of the uncertainties in weak decay inclusion into the proton yield. Though the parameter set has reproduced the ${^3_\Lambda}\mathrm{H}/{^3}\mathrm{He}$ ratio in LHC energy, the prediction has a slight down-shift at RHIC 200 GeV. 
 
Alike the $C_2$, this $S_3$ is important to relate the light nuclei and hypernuclei states to their composing nucleons and hyperons. I have estimated $S_3$ with and without decay contribution in $\Lambda$ and proton and have shown the variation in Fig.[\ref{fg.S3}]. First, I shall discuss the case with the decay feed-downs and check whether it can explain the available data or not, then shall follow up without the decay and check the similarity between the ratios $\Lambda/\mathrm{p}$ and ${^3_\Lambda}\mathrm{H}/{^3}\mathrm{He}$.  

With the decay feed-down, the phase space occupancy factor increases from 0.6 (AGS value) to $1$ at RHIC 200 GeV, and it drops to 0.6 at LHC.  $S_3$ remains flat near 0.6 SPS energies, which was previously shown by Ref.\cite{Andronic:2010qu}. Available data from experimental collaborations also support this non-monotonic behavior. The prediction for AGS energy is within the uncertainty band of data. The variation with collision energies arises due to the difference in decay contribution from hyperons and non-strange baryonic resonances. Contrarily, when one considers only the primary yields of $\Lambda$ and $\mathrm{p}$, the thermal model prediction for $S_3$ stays near $0.9$ at all $\sqrt{s_{NN}}$. 

As I have suitably reproduced the experimental data of $S_3$ with the total yields, the result regarding the primary density will be a guideline to investigate the $\Lambda$-hypernuclei formation. If the nuclei and hypernuclei formation occur near the hadronic chemical freeze-out and before the feed-down into $\Lambda$ and proton takes place, then there will be no significant differences between the $primary$ $\Lambda/p$ and ${^3_\Lambda}\mathrm{H}/{^3}\mathrm{He}$. In that case, the $S_3$ will stay near $1$ at all $\sqrt{s_{NN}}$. I have observed this flatness of $S_3$ in the thermal model predictions. This close resemblance between primary $\Lambda/p$ and ${^3_\Lambda}\mathrm{H}/{^3}\mathrm{He}$ indicates that the hypernuclei formation occurs from the primordial nuclei and hyperons.   

\subsection{Tritium to $^3 \mathrm{He}$ ratio}
The ratio of particles related to the same isospin multiplet helps to understand the isospin variation in the heavy-ion collision. In this context, the neutron to proton and $\pi^-/\pi^+$ are the representatives of the isospin asymmetry at the hadronic sector. The detected spectra of the neutron are not available in most of the $\sqrt{s_{NN}}$, so the ratio of $\pi^-$ and $\pi^+$ represents the variation of net isospin. The neutron to proton ratio remains $1.5$ in the colliding heavy-ions (Pb or Au). The initial isospin asymmetry generates net negative isospin in the final spectra, which increases at the lower $\sqrt{s_{NN}}$. Net negative isospin will favor an abundance of $\pi^-$ than its antiparticle. This effect will decrease at higher RHIC and LHC energies and the ratio $\pi^-/\pi^+$ becomes $1$. The ratio $^3 \mathrm{H}/ ^3 \mathrm{He}$ represents the isospin asymmetry in the light nuclei sector. Tritium $(^3  \mathrm{H})$ is composed of n-n-p, whereas $^3 \mathrm{He}$ is a n-p-p bound state. Therefore the tritium ($^3 \mathrm{H}$) to $^3 \mathrm{He}$ ratio should reveal the neutron to proton ratio \cite{Anticic:2016ckv}. 

\begin{figure}[!htb]
\subfloat[]{
{\includegraphics[scale=0.85]{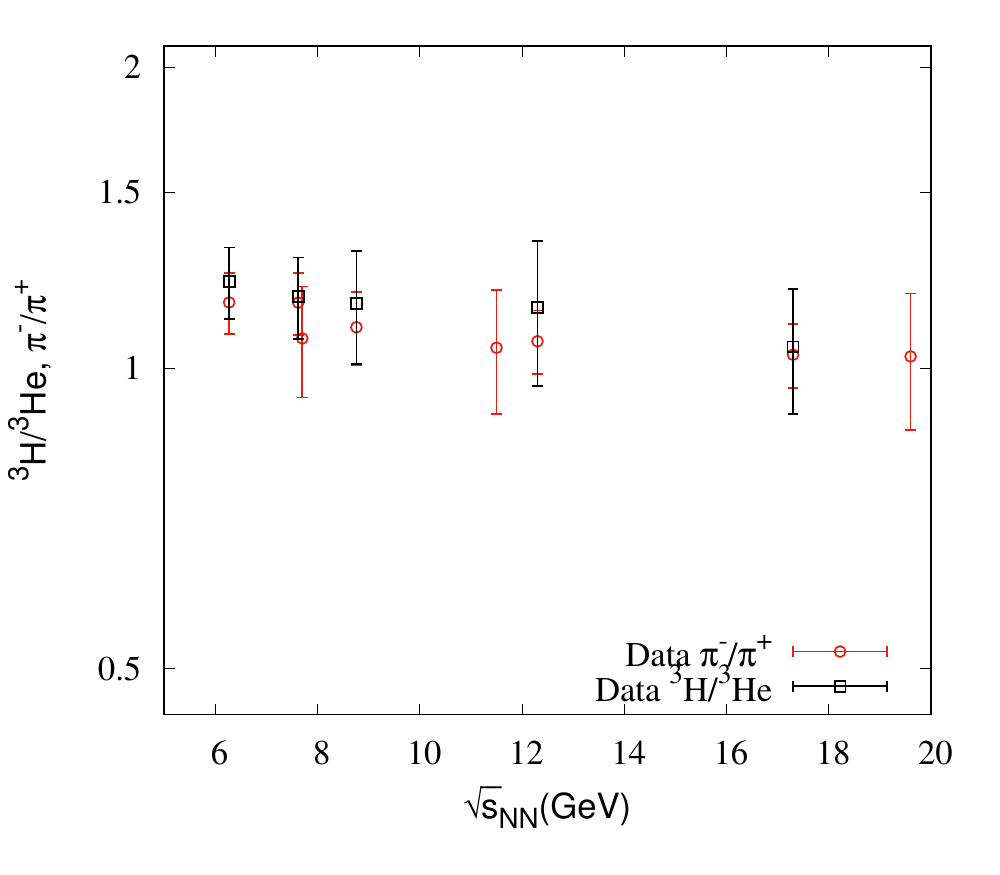}}
\label{fg.the3exp}
}\\
\subfloat[]{
{\includegraphics[scale=0.85]{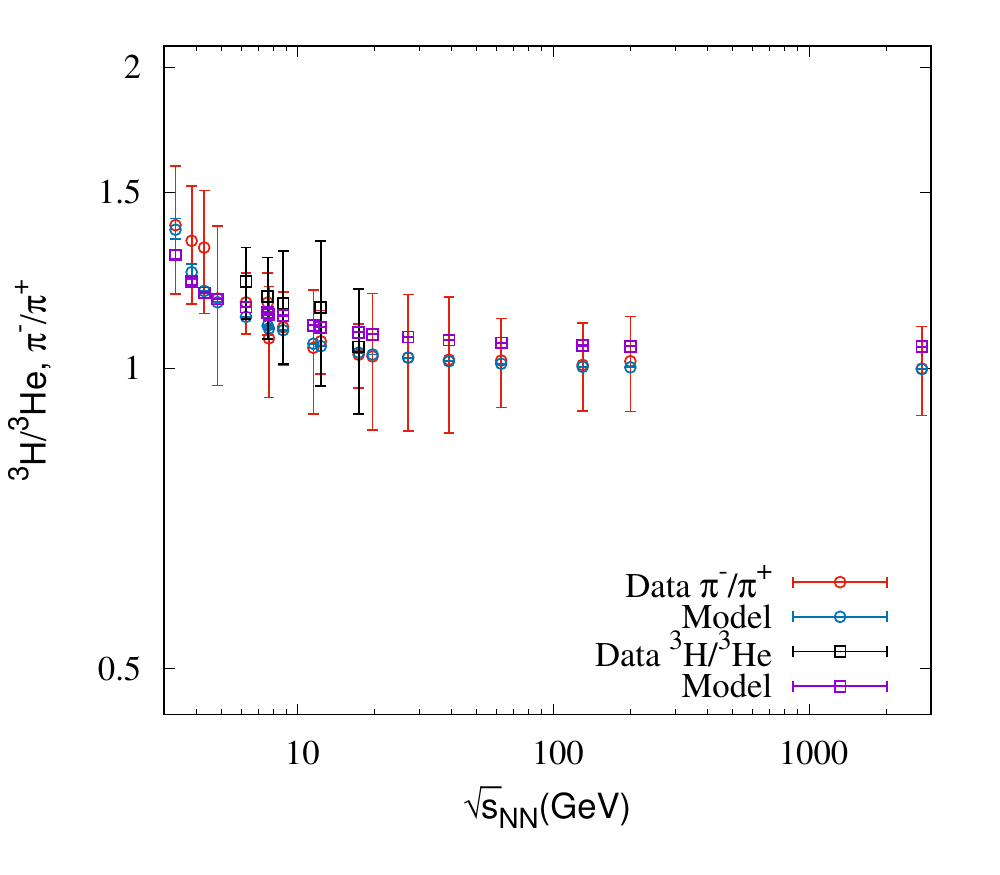}}
\label{fg.the3model}
}
\caption{
\label{fg.the3}
(a) Experimental data of $\pi^-/\pi^+$ (red) and $^3 \mathrm{H}/^3 \mathrm{He}$ (black) in SPS \cite{Anticic:2016ckv}. (b) Variation of thermal model predictions for $\pi^-/\pi^+$ (blue) and $^3 \mathrm{H}/^3 \mathrm{He}$  (violet). }
\end{figure}

The experimental data for tritium to $^3 \mathrm{He}$ is available in SPS energy\cite{Anticic:2016ckv}. Fig.[\ref{fg.the3exp}] displays data of $\pi^-/\pi^+$ ratio alongside tritium to Helium-3. The $^3 \mathrm{H}/^3 \mathrm{He}$ ratio has a close similarity with the pion ratio. The tritium and $^3 \mathrm{He}$ differ only in isospin and charge, like the charged pions. So the isospin asymmetry of the thermal source should be observed in $^3 \mathrm{H}/^3 \mathrm{He}$. 

In a thermal model, this isospin asymmetry generates a non-zero value of the corresponding chemical composition ($\mu_I$). Considering the Gell-Mann–Nishijima relation, one can use $\mu_Q$ instead of $\mu_I$. In Fig.[\ref{fg.the3model}], the model predictions for both $\pi^-/\pi^+$ and $^3 \mathrm{H}/^3 \mathrm{He}$ have been plotted. The $\mu_Q$ guides the $\sqrt{s_{NN}}$ variation of these ratios. The neutron and proton asymmetry of the colliding nuclei will dynamically propagate in the final state and induce an abundance of hadrons and nuclei with negative isospin value. Baryon stopping amplifies this asymmetry via large nucleon deposition in lower $\sqrt{s_{NN}}$ and increases these ratios. It is indeed interesting to observe that both the ratio $\pi^-/\pi^+$ and $^3 \mathrm{H}/^3 \mathrm{He}$ resemble each other, though their respective masses are widely different. This behavior proposes that the light nuclei share the same chemical freeze-out surface with that of the hadrons.

Here, I want to mention that, the double ratio $N_t N_p/{N{_d}{^2}}$ from the thermal model will be important in this context. But individual yields for tritium (t) yields in all the relevant experiments are still preliminary (HADES, STAR, ALICE).

\section{Summary and outlook}\label{sec4}
The description of light nuclei in a thermal model holds difficulties due to their small binding energy. Various studies within the framework of the thermal model have addressed these uncertainties of the weakly bound states at the chemical freeze-out surface \cite{Andronic:2017pug, Oliinychenko:2018ugs, Cai:2019jtk}. Despite the possible inconsistencies, the Statistical Hadronization Model (SHM) provides a reasonable description of the yields of light nuclei and hypernuclei. In this manuscript, I have revisited the light nuclei equilibration at the chemical freeze-out of the heavy-ion collision. I have performed the parametrization with ratios of the net baryon charge to total baryon charge and total multiplicity and have verified the efficiency of the parameter set by comparing thermal model predictions with available experimental data.

Separate ratios have been addressed to check the light nuclei equilibration in the baryon, strangeness, and isospin sector. The light nuclei to proton ratios have been employed to discuss the equilibrium in the baryon sector. On the other hand, a proper agreement between the thermal model and data for the ratio ${^3_\Lambda}\mathrm{H}/{^3}\mathrm{He}$ signifies the strangeness-baryon equilibrium in light nuclei. In the context of isospin, one finds resemblance between the ratios ${^3}\mathrm{H}/{^3 \mathrm{He}}$ and $\pi^-/\pi^+$. Both of these ratios carry the information of isospin asymmetry, within a thermal model prescription.

An essential outcome of the present work is a proper thermal model description of the strangeness population factor $S_3$. I have found a good agreement with data at both RHIC-200 and LHC-2.76 TeV. The equilibrium in the hypernuclei sector is apparent from the agreement between the thermal model and data. The successful description from the thermal model emphasizes the fact that the light nuclei exist in equilibrium with the hadrons at the chemical freeze-out boundary. 

This study has especially examined the relationship between the light nuclei ratios and their hadronic counterpart $\mathrm{\bar{d}/d}$, $(\mathrm{\bar{p}/p})^2$ and $\Lambda/p$, ${^3_\Lambda}\mathrm{H}/{^3}\mathrm{He}$ to discuss the formation and freeze-out of the light nuclei and hypernuclei. First, I have reviewed the individual ratios with the standard thermal model prescription. Then I have proposed that a better resemblance between light nuclei, hypernuclei ratios and their hadronic counterpart can be found without the decay contribution in the final yields of hadrons. These results denote that the formation of light nuclei and hypernuclei takes place long before the decay of resonances occurs to constituting hadrons. In that case, the ratio of the primordial yields of the hadronic constituents is a good estimation of the light nuclei and hypernuclei ratios. 

At this juncture, it should be mentioned that a complete description of light nuclei and hypernuclei yields demands a framework considering both the dynamical plus chemical interaction among hadrons and light (hyper)nuclei states. Various possibilities may produce an equilibrium thermal description of these (hyper)nuclei states. They may be born near the chemical freeze-out and evolve without further interaction, or they can maintain the equilibrium abundances via successive creation, annihilation, and regeneration. This investigation is beyond the scope of the present framework of a dilute gas of hadrons, nuclei, and resonances \cite{Cai:2019jtk}. In this context, Ref.\cite{Oliinychenko:2018ugs} has discussed the deuteron yields at LHC in a hybrid approach and shown that the deuterons may still chemically interact after the chemical freeze-out, but their yields do not vary much from the equilibrium thermal value. It is a subject of future investigation whether this picture holds for other nuclei and hypernuclei or not, but this approach opens up a possible scenario for the equilibrated yields of nuclei and hypernuclei. Despite these uncertainties regarding the existence of these weakly bound states, the successful thermal model description with a minimal number of parameters advocates that the light nuclei and hypernuclei are in thermal equilibrium at the chemical freeze-out surface.  

To summarize, this study introduces a new approach to investigate the relationship among the light nuclei to their hadronic constituents at freeze-out. By turning on and off the decay feed-down in the hadrons, I have shown that a better correlation between light nuclei ratios and corresponding hadronic ones can be found when the decay feed-down into hadrons are excluded. These results indicate that the light nuclei ratios become fixed near the standard chemical freeze-out surface and before the decay of the hadronic resonance occurs. This method will serve as a benchmark to discuss the formation of light nuclei and the inclusion of decay into their constituents.  This method is applicable only for the ratio of mass clusters with the same mass number. I shall address this issue with other light nuclei and hypernuclei yields from the expected results from the RHIC-BES and SPS, CBM at FAIR. 

\section*{ACKNOWLEDGEMENTS}
This work is funded by UGC and DST of the Government of India. I take this opportunity to thank Sumana Bhattacharyya, Dr. Sanjay K. Ghosh, and Dr. Rajarshi Ray for the various discussion regarding the statistical thermal Model. Author thanks Samapan Bhadury and Pratik Ghosal for the critical reading of the manuscript.

\bibliography{refcentral}

\end{document}